\documentclass[aps,prl,twocolumn,groupedaddress,floatfix]{revtex4}

\usepackage{graphicx}
\usepackage{amssymb}

\begin{document}

\title{Noise, coherent fluctuations, and the onset of order in an oscillated granular fluid}
\author{Daniel I. Goldman}
\email[]{goldman@chaos.utexas.edu}
\homepage{http://chaos.ph.utexas.edu/~goldman}
\author{J. B. Swift}
\author{Harry L. Swinney}
\affiliation{Center for Nonlinear Dynamics,
                The University of Texas at Austin,
                Austin, TX 78712}
\date{\today}

\begin{abstract}

We study fluctuations in a vertically oscillated layer of grains below the critical acceleration for the onset of ordered standing waves. As onset is approached, transient disordered waves with a characteristic length scale emerge and increase in power and coherence. The scaling behavior and the shift in the onset of order agrees with the Swift-Hohenberg theory for convection in fluids. However, the noise in the granular system is four orders of magnitude larger than the thermal noise in a convecting fluid, making the effect of granular noise observable even $20\%$ below the onset of long range order.

\end{abstract}


\maketitle

Recent experiments~\cite{exprefs} and simulations~\cite{simrefs} on flows of dissipative granular particles have been found to be described by hydrodynamic theory, although the granular systems exhibit much larger fluctuations than fluids -- a single wavelength in a pattern in a vibrated granular layer might contain only $10^2$ particles instead of the $10^{22}$ particles in one wavelength of a pattern in a vibrated liquid layer. Thus fluctuations play a more significant role in granular systems than in fluid flows. In fluids the fluctuations are driven by thermal noise and are described by the addition of terms to the Navier-Stokes equations~\cite{landau,zaiAshl}. This fluctuating hydrodynamic theory has been found to describe accurately the dynamics  near the onset of a convection pattern in a fluid ~\cite{wuAahl,rehAras,swiAhoh}. Here we describe an experimental study of fluctuations near the onset of square patterns in granular layers, and we demonstrate the applicability of fluctuating hydrodynamics to this inherently noisy system.

\begin{figure}
\begin{center}
\includegraphics[width=3.25in]{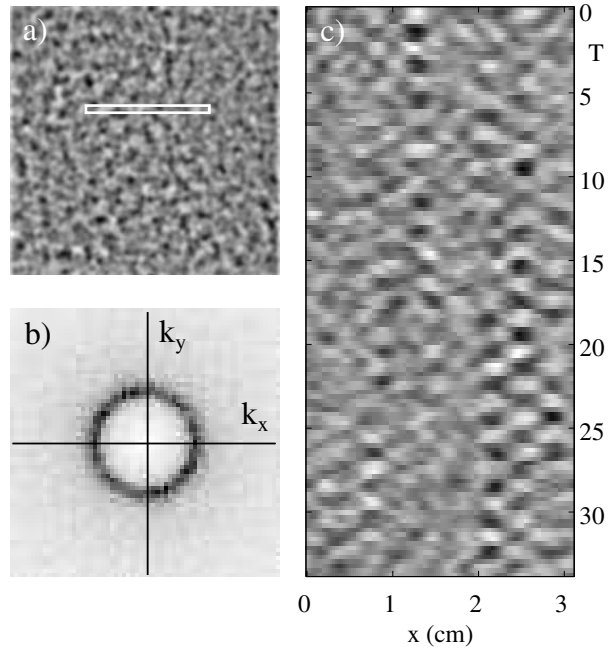}

\caption{\label{figure1} (a) Snapshot of an area $6.25 \times
6.25~\mbox{cm}^2$ in a container oscillating with $\Gamma=2.6$.
(b) The spatial power spectrum of (a) has an intense ring
corresponding to randomly oriented spatial structures with a
length scale of 0.52 cm (for improved signal-to-noise, 100 spectra
were averaged to obtain the spectrum shown). (c) Space-time
diagram for the row of pixels in the box in (a); the period of the
localized transient oscillations is $2T\equiv 2/f_d$.}

\end{center}
\end{figure}

{\em Experiment ---} We study a vertically oscillating layer of
$170~\mu$m stainless steel particles as a function of $\Gamma$,
the peak plate acceleration relative to gravity; the container oscillation frequency $f_d$ is fixed at 30
Hz~\cite{melAumb}. The layer (depth 5 particle diameters)
fluidizes at $\Gamma \approx 2$~\cite{mujAmel}, and a square wave
pattern with long range order emerges at $\Gamma \approx 2.77$.
The layer is illuminated at a low angle and fluctuations in the
surface density are observed as fluctuations in the light
intensity~\cite{filter1revised}, recorded on a 256$\times$256
pixel CCD camera.

\begin{figure}[h!tb]
\begin{center}
\includegraphics[width=3in]{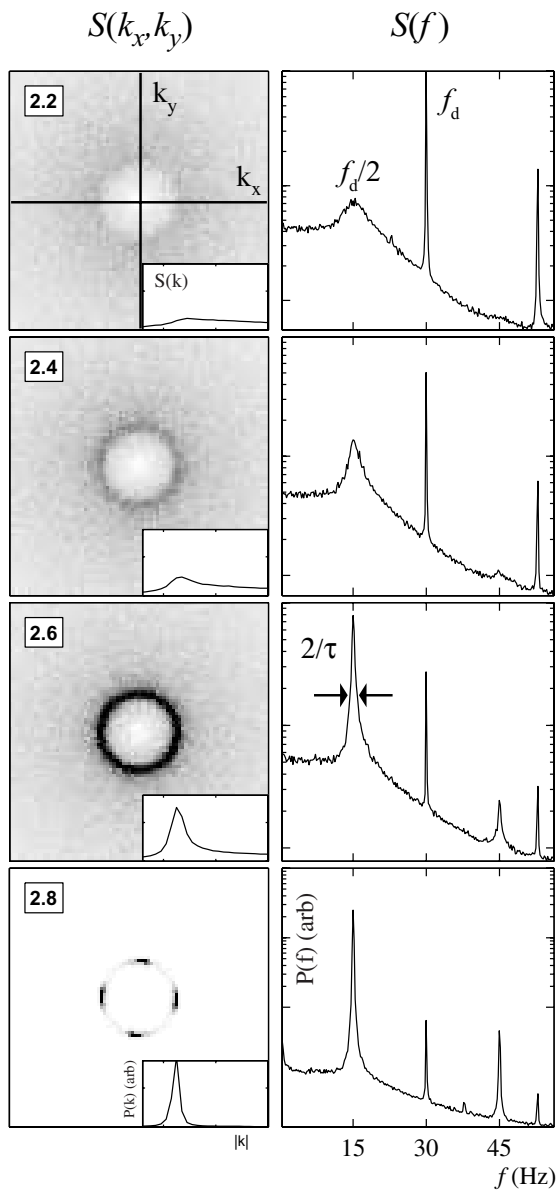}
\caption{\label{figure2} Spatial power spectra $S(k_x,k_y)$ and
temporal spectra $S(f)$ and structure factors $S(k)$ (insets) show
that the fluctuations increase in power and coherence as $\Gamma$
increases from 2.2 to 2.6, and at $\Gamma=2.8$, above onset, a
coherent square spatial pattern has emerged. $S(k_x,k_y)$, $S(k)$,
and $S(f)$ have been divided by $10$ at $\Gamma=2.8$, where there
is a strong peak at 15 Hz corresponding to the spatial pattern
(this peak mixes with the 30 Hz drive frequency to give a peak at
45 Hz). The value of $k$ corresponding to a maximum in the structure function for $\Gamma=2.8$ corresponds to a length of 0.52 cm.}
\end{center}
\end{figure}

{\em Coherent fluctuations ---} Fluctuations are evident in the snapshot of the layer shown in Fig.~\ref{figure1}(a). Spatial coherence with no orientational order is revealed by the ring in the spatial power spectrum $S(k_x,k_y)$ shown in Fig.~\ref{figure1}(b). The increase in the spatial and temporal coherence and the power of the fluctuations with increasing $\Gamma$ is illustrated in Fig.~\ref{figure2}, where insets with each spectrum $S(k_x,k_y)$ show the corresponding azimuthally averaged structure factor, $S(k)={\langle S(k_x,k_y) \rangle}_\theta$ with $k=\sqrt{k_x^2+k_y^2}$. (There is slight ($<1 \%$) hysteresis in all measured quantities between increasing and decreasing $\Gamma$, but we will only discuss increasing $\Gamma$.) The power of the dominant mode increases while its width decreases with increasing $\Gamma$~\cite{wavevec}. The noise is readily visible at $\Gamma=2.2$, which is 20\% below the onset of long range order, while in high sensitivity experiments on Rayleigh-B\'{e}nard convection, the noise became measurable only at $1\%$ below the onset of convection~\cite{wuAahl}.

{\em Local transient waves ---}  Transient localized structures
oscillating at $f_d/2$ are visible in Fig.~\ref{figure1}(c).
Temporal power spectra of the intensity time series for each pixel
in the images are averaged to obtain the power spectra $S(f)$
shown in Fig.~\ref{figure2}. The peak at $f_d/2$ emerges for
$\Gamma \gtrsim 2$ and increases in power with increasing $\Gamma$. The
half-width at half-maximum of the peak, denoted $1/\tau$,
decreases as $\Gamma$ increases, indicating increasing temporal
coherence of the noisy structures. Except for the peak at $f_d/2$,
$S(f)$ does not depend strongly on $\Gamma$ for $\Gamma \geq 2$. The
shape of $S(f)$ is similar to that for surface excitations
(ripplons) on a fluid driven by thermal noise~\cite{katAing}.

%%%%%%%%% From here we compare to fluc. hydro theory %%%%%%%%%

{\em Fluctuating hydrodynamics ---} The phenomena we have described have the features of noise-driven damped hydrodynamic modes close to a bifurcation. We now interpret the observations using the Swift-Hohenberg (SH) model, which is based on the Navier-Stokes equation and was developed to describe noise near the onset of long range order in Rayleigh-B\'{e}nard convection~\cite{swiAhoh}. Swift-Hohenberg theory predicts that below the onset of ordered patterns, noise drives a ring of modes that increases in power as onset is approached; our observations in Fig.~\ref{figure2} are in qualitative accord with this prediction. The theory also predicts that the nonlinearity of the fluid acting on the noise will lead to an increase of the critical value of the bifurcation parameter for the onset of long range order.  The observations also agree qualitatively with this prediction: for the stainless steel particles the patterns are noisier than those obtained in previous experiments on lead particles, which are more dissipative; further, the onset of long range order for the stainless steel particles occurs for $\Gamma \approx 2.77$, while for lead, $\Gamma \approx 2.5$~\cite{bizAsha}.

The SH model describes the evolution of a spatial scalar field
$\psi({\bf x},t)$,

\begin{equation}
\label{sheq}
 \frac{\partial \psi}{\partial t} = \bigl(\epsilon-(\nabla^2+k_0^2)^2
 \bigr)\psi-\psi^3 + \eta({\bf x},t),
\end{equation}

\noindent where $\epsilon$ is the bifurcation parameter and $\eta$ is a stochastic noise term such that $\langle \eta({\bf x},t) \eta({\bf x'},t') \rangle = 2 F \delta({\bf x}-{\bf x'})\delta(t-t')$, where $F$ denotes the strength of the noise. In the absence of noise ($F=0$), called the mean field (MF) approximation~\cite{swiAhoh,schAahl}, there is an onset of stripe patterns with long-range order at $\epsilon=\epsilon_c^{\mbox{\tiny MF}}=0$. (Our experiments yield squares at pattern onset with slight hysteresis, but we compare our observations to the the simplest model for noise interacting with a bifurcation, Eq.~\ref{sheq}, which yields stripes at onset via a forward bifurcation; a more complicated model yielding square patterns and hysteresis is described in~\cite{sakAbra}.) For $F \neq 0$, the onset of long-range (LR) order is delayed until $\epsilon=\epsilon_c^{\mbox{\tiny LR}}>0$, where $\epsilon_c^{\mbox{\tiny LR}} \propto F^{2/3}$.  For $0<\epsilon<\epsilon_c^{\mbox{\tiny LR}}$, the pattern is disordered, and appears cellular-like~\cite{xiAvin}. We define the delay in onset as $\Delta \epsilon_c=\epsilon_c^{\mbox{\tiny LR}}-\epsilon_c^{\mbox{\tiny MF}}$.

\begin{figure}[h!tb]
\begin{center} 
\includegraphics[width=3in]{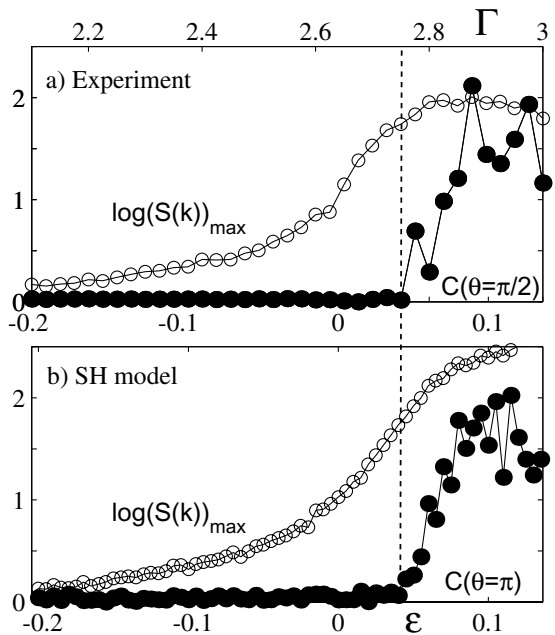}
\caption{\label{figure4} The growth of the noise power and the onset of long-range order in (a) experiment and (b) Swift-Hohenberg model. The log of the maximum of $S(k)$ ($\circ$) increases through the mean field onset ($\epsilon_c^{\mbox{\tiny MF}}=0$), while the onset of long range order, indicated by appearance of angular correlations of the radially averaged structure factor [$C(\theta=\pi/2)$ for the experiment and $C(\theta=\pi)$ for SH equation ($\bullet$)], is delayed to $\epsilon_c^{\mbox{\tiny LR}} \approx 0.04$. The integration of Eq.~\ref{sheq} uses a scheme described in~\cite{croAmei94}; the solution is obtained on a $128 \times 128$ grid with $k_0=1$ and integration time-step 0.5.} 
\end{center}
\end{figure}

To compare results from the experiments with results obtained by integration of the SH model, we must first compute the reduced control parameter $\epsilon=(\Gamma-\Gamma_c^{\mbox{\tiny MF}})/\Gamma_c^{\mbox{\tiny MF}}$. However, we have no {\it a priori} way to determine $\Gamma_c^{\mbox{\tiny MF}}$ since the theory predicts a smooth change in all quantities as the mean field bifurcation is crossed. We determine $\Gamma_c^{\mbox{\tiny MF}}$ by performing a two-parameter fit to obtain agreement between experiment and the SH model for the maximum value of $S(k)$ for all $\epsilon$ below onset and for the critical value for the emergence of long range order $\epsilon_c^{\mbox{\tiny LR}}$ as determined from the onset of angular correlation in the radially averaged structure factor ($C(\theta=\pi/2)$ for the emergence of squares in the experiments and $C(\theta=\pi)$ for stripes in the SH model). In the fitting procedure we vary $\Gamma_c^{\mbox{\tiny MF}}$ for the experimental data and $F$ in Eq.~\ref{sheq}~\cite{fitnote}. Thus, in addition to a prediction of $\Gamma_c^{\mbox{\tiny MF}}$, the procedure allows us to estimate the size of the noise required to shift $\epsilon_c^{\mbox{\tiny MF}}=0$ up to the observed $\epsilon_c^{\mbox{\tiny LR}}$, i.e., to shift $\Gamma_c^{\mbox{\tiny MF}}$ up to $\Gamma_c^{\mbox{\tiny LR}}= 2.77$.  

The two-parameter fit yields $\Gamma_c^{\mbox{\tiny MF}} = 2.64$ and $F=3.5 \times 10^{-3}$, giving a delay in onset of $\Delta \epsilon_c=0.04$ (Fig.~\ref{figure4}). The noise strength is a factor of $10^4$ larger than the noise in the convecting fluid experiments of Wu {\it et al.}~\cite{wuAahl} and is a factor of $10$ larger than in recent experiments on convection near a critical point~\cite{ohAahl1}. The theory also predicts that at $\epsilon_c^{\mbox{\tiny LR}}$ there should be a jump in $\langle \psi^2 \rangle$ proportional to $F^{2/3}$; however, the predicted jump, only $10^{-3}$ in $\langle \psi^2 \rangle$, is too small to detect with the precision of our experimental measurements. (Also, such an effect would be dominated by the hysteresis in the transition to square patterns.) Now that we know the appropriate reduced control parameter $\epsilon$ for the experiment, we can test predictions of scaling given by Eq.~\ref{sheq}.

\begin{figure}[h!tb]
\begin{center}
\includegraphics[width=2.3in]{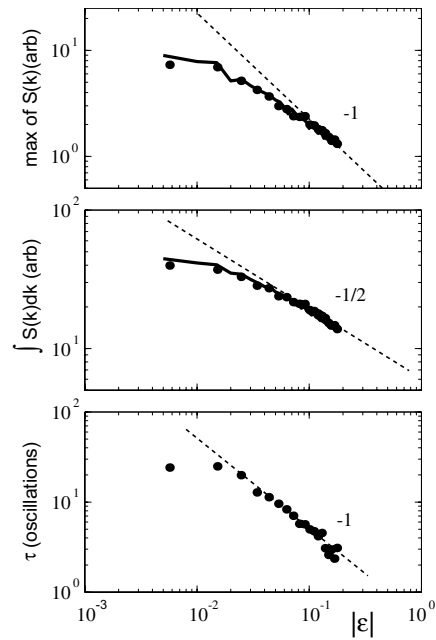}
\caption{\label{figure5} The noise peak intensity, total noise power, and the noise coherence time as a function of the distance $\epsilon$ below the onset of patterns. The measured quantities ($\bullet$) compare well to the Swift-Hohenberg equation (solid curve) with $F=3.5 \times 10^{-3}$. Results from both experiment and SH theory approach the predictions from the linear theory (shown by the dashed lines with the values for their slopes) when $\epsilon$ is large, far below the onset of long range order.}
\end{center}
\end{figure}

{\em Scaling ---} Linear theory predicts the following scalings for thermal convection with {\em small} noise levels near the onset of long range order: both the noise peak intensity (at the wavenumber selected by the system) and the correlation time should scale as $|\epsilon|^{-1}$ and the power in the fluctuations should scale as $|\epsilon|^{-\frac{1}{2}}$~\cite{wuAahl,zaiAshl}. The scaling behavior found in both the experiment and SH model for large noise levels is shown in Fig.~\ref{figure5}, where the noise peak intensity was determined from the maximum of $S(k)$ (see insets of Fig.~\ref{figure2}), and the power in the fluctuations was determined from the area under the peak in $S(k)$, after subtracting the approximately constant background due presumably to incoherent grain noise. The Swift-Hohenberg theory agrees remarkably well with the observations.  As expected, both experiment and theory deviate considerably from the linear theory prediction for small $|\epsilon|$, where nonlinear effects are large, but for large $|\epsilon|$ the experiment and simulation approach the scaling predicted by the linear theory. Finally, we have determined from the experimental data the correlation time for the patterns by fitting the $f_d/2$ peak of $S(f)$ to a Lorentzian and computing the half-width at half maximum, $1/\tau$. Again, for large $\epsilon$ the observations approach the scaling predicted by linear theory, but deviate from linear theory for small $\epsilon$ (Fig.~\ref{figure5}).

%%%%%%%%%%%%%%%%% Conclusions %%%%%%%%%%%%%%%%%%%%%%%%%%

{\em Conclusions ---} We have shown that a vertically oscillated layer of grains exhibits behavior consistent with the theory of fluctuating hydrodynamics for Navier-Stokes fluids. This indicates that fluctuations must be included in the hydrodynamic equations for granular media~\cite{meeApos}. In fact, the large fluctuations present in granular fluids suggests that such systems could be useful to study the effects of noise in nonequilibrium fluids {\it far} below a bifurcation~\cite{zarAsen,ohAahl2}.

Like numerical studies of elastic gases~\cite{manAgar}, our experiments and simulations show that fluctuating hydrodynamics can apply down to length scales of only a few mean free paths.  Fluctuations are important in fluids at the nanoscale, which are of current interest~\cite{eggers02,mosAlan}. The fluctuations are difficult to study in gases and liquids but can be studied easily in granular materials, which may demonstrate some essential features of the nanoscale flows.

\begin{acknowledgments}
We thank Guenter Ahlers for helpful discussions and for providing preliminary data. This work was supported by the Engineering Research Program of the Office of Basic Energy Sciences of the U. S. Department of Energy (Grant No. DE-FG03-93ER14312), The Texas Advanced Research Program (Grant No. ARP-055-2001), and the Office of Naval Research Quantum Optics Initiative (Grant N00014-03-1-0639).
\end{acknowledgments}

%% Note: For bibliography, first compile with noisebib, then cut and paste from .bbl file and modify Vinals and Xi, and cut dates from condmat.

%\bibliography{f:/sciwork/papers/wavenoise/submit1/noisebib}

\begin{thebibliography}{25}
\bibitem[{fil()}]{filter1revised}
\bibinfo{note}{To measure the fluctuations in the intensity, we subtract from
  each measurement the average of 100 images taken at the same phase of the
  driving cycle and then divide each difference by the average image, as
  in~\cite{wuAahl}.}
\bibitem[{wav()}]{wavevec}
\bibinfo{note}{As $\Gamma$ is increased from $2.0$ to $2.8g$, the wavevector of
  the dominant mode of the fluctuations smoothly decreases by about a factor of
  1.5 as more of the depth of the layer becomes fluidized.}
\bibitem[{fit()}]{fitnote}
\bibinfo{note}{Once the fit parameters are determined, $S(k)$ in the SH model
  is scaled to correspond to $S(k)$ in the experiment.}
\end{thebibliography}

\end{document}